\documentclass[aps,prb,showpacs,preprintnumbers,twocolumn,superscriptaddress]{revtex4-2}
\usepackage{amsmath,amssymb}
\usepackage{bm}
\usepackage{tipa}
\usepackage{upgreek}
\usepackage{comment}
\usepackage{mathrsfs}
\usepackage{graphicx}
\usepackage{multirow} 
\usepackage{braket}
\usepackage{enumitem}
\usepackage{mathbbol}
\usepackage{booktabs}
\usepackage{gensymb}
\usepackage[normalem]{ulem}
\usepackage{color, xcolor}
\usepackage[colorlinks,bookmarks=true,citecolor=blue,linkcolor=red,urlcolor=blue]{hyperref}
\usepackage{hyperref}
\renewcommand{\vec}[1]{\mathbf{#1}}
\usepackage{pifont}
\newcommand{\cmark}{\ding{51}}
\newcommand{\xmark}{\ding{55}}
\newcommand{\bk}{\bm{k}}

\renewcommand{\comment}[1]{}
\usepackage{subfigure}
\newcommand\titlelowercase[1]{\texorpdfstring{\lowercase{#1}}{#1}}

\begin{document}

%\title{Non-Abelian states vs phase separation in opposite Chern number Landau levels}

\title{Fractional topological insulators at odd-integer filling: \\
Phase diagram of two-valley quantum Hall model}

\author{Sahana Das}
\affiliation{Department of Physics, University of Z{\"u}rich, Winterthurerstrasse 190, 8057 Z{\"u}rich, Switzerland}
\author{Glenn Wagner}
\affiliation{Institute for Theoretical Physics, ETH Z{\"u}rich, 8093 Z{\"u}rich, Switzerland}
\author{Titus Neupert}
\affiliation{Department of Physics, University of Z{\"u}rich, Winterthurerstrasse 190, 8057 Z{\"u}rich, Switzerland}

\begin{abstract}
The fractional quantum Hall effect has recently been shown to exist in heterostructures of van der Waals materials without an externally applied magnetic field, e.g. in twisted bilayers of MoTe$_2$. These fractional Chern insulators break time-reversal symmetry spontaneously through polarization of the electron spins in a quantum spin Hall insulator band structure with flat bands. This prompts the question, which states could be realized if the spins remain unpolarized or polarize partially. Specifically, the possibility of time-reversal symmetric topological order arises. Here, we study this problem for odd integer filling of the bands, specifically focusing on vanishing and half valley polarization. Short of reliable microscopic models for small twist angles around $2.1^\circ$, we study the idealized situation of two Landau levels with opposite chirality, the two-valley quantum Hall model. Using exact diagonalization, we identify different phases arising in this model by tuning the interaction. In the physically relevant regime, the system initially exhibits phase-separated or valley-polarized states, which eventually transition into paired states by reducing onsite Coulomb repulsion.

\end{abstract}

\maketitle

\section{Introduction}

Moiré materials have led to a slew of discoveries of topological and correlated phases \cite{Andrei2021}. The twist angle in moiré homobilayers serves as a tuning knob that can be used for band structure and interaction engineering. In twisted transition metal dichalcogenides, lowering the twist angle increases the ratio of interaction energy to bandwidth \cite{Reddy2023}, such that the low twist angle regime allows access to strong correlations. Furthermore, the bands have a topological character manifesting in non-zero Chern numbers. In particular, in twisted MoTe$_2$ at twist angles of around $2.1^\circ$, the band structure closest to zero doping consists of a pair of spin-valley locked bands with opposite valleys having opposite Chern numbers \cite{Wu2019}. Gating the device allows for doping into these flat bands. 

A fractional Chern insulator \cite{neupert,sheng,regnault} is the zero-field analogue of the venerable fractional quantum Hall effect. Due to its similarity with a Landau level, a nearly flat Chern band \cite{Tang2011,Sun2011} in the presence of strong correlations is the ideal setting for realizing a fractional Chern insulator. Indeed, both optics and transport experiments in twisted MoTe$_2$ around $2.1^\circ$ have proven the existence of spontaneously spin-polarized fractional Chern insulators at commensurate filling factors $\nu=-2/3,-3/5,\dots$, corresponding to the Jain sequence of fractional quantum Hall states \cite{cai2023signatures,zeng2023integer,Park2023,Xu2023,park2024ferromagnetism,xu2024interplay}. The Jain sequence of fractional quantum Hall states can be understood as integer quantum Hall states of composite fermions, bound states of an electron and an even number of flux vortices that experience a reduced effective magnetic field~\cite{Lopez1991}. At filling factor $\nu=-1/2$ ($\nu=-1/4$), the composite fermions consisting of an electron bound to two (four) flux vortices experience no magnetic field and hence form a compressible composite Fermi liquid (CFL) \cite{HLR}. The CFL phase at $\nu=-1/2$ has been obtained in exact diagonalization studies of twisted MoTe$_2$ \cite{Dong2023,Goldman2023} and recent experimental evidence has corroborated this finding \cite{Anderson2024}.

While twisted MoTe$_2$ around $2.1^\circ$ has similarities to Landau level physics, there is a striking difference. In a given spin sector, the two bands closest to zero doping have opposite Chern numbers \cite{cai2023signatures,zeng2023integer,Park2023,Xu2023}. The second-lowest band hence does not serve as an analogue of the second Landau level (SLL), which is known to host non-Abelian fractional quantum Hall phases. Furthermore, even when one is doping into the analogue of the lowest Landau level (LLL), the effect of band mixing and the consequence of significant band dispersion will differentiate the system strongly from one with Landau levels in the quantum Hall effect. This motivates the search for a better Landau level analogue, which is found in twisted MoTe$_2$ around $2.1^\circ$. In this regime, the three lowest bands in a given spin sector have the same Chern number ($\pm 1$) and the quantum geometry of the bands is very similar to that of a Landau level sequence \cite{Ahn_2024,Wang2025,Xu2025,liu2024generalizedlandau,Li2025,Wang2023,Morales2024,Shi2024}.

Experiments on twisted MoTe$_2$ around $2.1^\circ$ have seen evidence of an insulating state at $\nu=-3$ \cite{kang2024evidence_nature,kang2025timereversalsymmetrybreakingfractional}. This state has a vanishing Hall conductance, i.e.~it cannot be spin-polarized integer Chern insulator. The nature of the state is still unexplained.  A natural candidate would be a fractional topological insulator, a time-reversal symmetric state which is the lattice analogue of the fractional quantum spin Hall effect \cite{bernevigQSHE2006,Levin_Stern,Neupert2011FTI,Stern2015review,Neupert_2015,Levin2012classification}, although this has been ruled out by the recent observation that time-reversal symmetry is broken in this state \cite{kang2025timereversalsymmetrybreakingfractional}. Some proposals consistent with time-reversal breaking are tensor products of Moore-Read (MR) states \cite{abouelkomsan2024nonabelian} that would occur at filling factor $\nu=\nu_\uparrow+\nu_\downarrow=-3/2-3/2$ in the second Landau level. Alternatively, a partially particle-hole transformed Halperin (331) state at filling $\nu=-5/4-7/4$ has been proposed \cite{Sodemann}. 

With a view towards understanding the phases that can arise in twisted MoTe$_2$ around $2.1^\circ$, we study a model with two Landau levels of opposite Chern numbers, the two-valley quantum Hall (TVQH) model \cite{Chen2012FQHtorusFTI,furukawa2014global,Repellin2014FTI,Mukherjee2019FQHsphereFTI,Bultinck2020mechanism,zhang2018composite,kwan2021exciton,kwan2022hierarchy,eugenio2020DMRG,stefanidis2020excitonic,chatterjee2022dmrg,myersonjain2023conjugate,yang2023phase,Wu2024,shi2024excitonic,kwan2024textured,abouelkomsan2023band,kwan2024abelianfractionaltopologicalinsulators, wagner2025variationalwavefunctions,brunner2025tuningfractionaltopologicalinsulator,zou2025valleyordermoiretopological,wang2025_CTI}. While the TVQH model can realize fractional topological insulators, it requires a suppression of the short-range intervalley Coulomb repulsion $V_0$ in order for them to be stabilized \cite{wagner2025variationalwavefunctions,abouelkomsan2024nonabelian,Sodemann,kwan2024abelianfractionaltopologicalinsulators,brunner2025tuningfractionaltopologicalinsulator}. Such a suppression of $V_0$ can be motivated by band-mixing in twisted MoTe$_2$ \cite{kwan2024abelianfractionaltopologicalinsulators,abouelkomsan2024nonabelianspinhallinsulator}, and could in general also arise from phononic contributions. In the absence of $V_0$ suppression, the fractional topological insulator gives way to phase separated (PS) or valley polarized states \cite{yang2023phase,Mukherjee2019FQHsphereFTI,Chen2012FQHtorusFTI,kwan2024abelianfractionaltopologicalinsulators}. 

We therefore focus on the phase diagram of the opposite Landau level model at total filling $\nu=1$ in both the LLL and SLL. We pay particular attention to the effect of $V_0$ suppression, known to be important for the magnetic properties \cite{kwan2024abelianfractionaltopologicalinsulators}. We work at filling factors $\nu=1/2+1/2$ and $\nu=1/4+3/4$. We start from the decoupled limit, where the inter-valley interaction is switched off. In that case, the separate valleys may either form a MR state or a CFL. At $\nu=1/2$ filling, the CFL is preferred in the LLL, while a MR is preferred in the SLL, at $\nu=1/4$ filling the CFL is preferred in both LLL and SLL \cite{Yutushui2025}.

\section{Methods}

In order to simulate a time-reversal symmetric transition metal dichalcogenide such as twisted MoTe$_2$, we consider the TVQH model where the Chern number $C=\tau$ is opposite for the two valleys $\tau=\pm1$ (microscopically the two valleys correspond to the $K$ and $K'$ valleys of the hexagonal Brillouin zone). We perform exact diagonalization on the TVQH model in the torus geometry. We denote by $N_e$ the total number of electrons and by $N_\phi$ the number of flux quanta penetrating the torus. The filling factor is
\begin{equation}
    \nu=N_e/N_\phi = p/q,
\end{equation}
where $p$ and $q$ are co-prime. Let $N_e=pN, N_\phi=qN$ where $N=\textrm{GCD}(N_e,N_\phi)$ and GCD stands for greatest common divisor.
\textcolor{black}{The torus is spanned by vectors $\Vec{L}_x=N_xa_M\hat{e}_x$ and $\Vec{L}_y=N_ya_M\hat{e}_y$, where $a_M=9.60$~nm is the moiré lattice constant for twisted MoTe$_2$ at a twist angle of $2.1^\circ$. We have $|\Vec{L}_x\times\Vec{L}_y|=L_xL_y\sin\theta=2\pi\ell_B^2N_\phi$, where $\theta$ is the angle between $\hat{e}_x$ and $\hat{e}_y$, while $\ell_B$ is the magnetic length. For the hexagonal Brillouin zone with $\theta=\pi/3$ we find $\ell_B\simeq 3.56$~nm.}

The many-body momenta take $N\times N_\phi$ values $K_x=0\cdots,N-1;\, K_y=0\cdots,N_\phi-1$ \cite{Bernevig2012}.
For the system with $\nu=p/q$, we have $q$-fold center of mass degeneracy. This results in the reduced number $N\times N$ of momenta $K_x=0\cdots,N-1;\, K_y=0\cdots,N-1$ \cite{Bernevig2012}. In our case, we always focus on the situation $N_e=N_\phi$ such that the total filling factor is $\nu=1$. We fix the number of electrons in the $\tau=+1$ valley, $N_+$, and the number in the $\tau=-1$ valley, $N_-$, such that the filling factors are either $\nu=\nu_++\nu_-=1/2+1/2$ or $\nu=1/4+3/4$.

We consider electrons interacting via a single-gate screened Coulomb interaction with gate distance $\xi=35$~nm.  The Fourier-transformed interaction potential takes the form
\begin{equation}
    V_\xi(q)=\frac{e^2}{2\epsilon}\frac{1-e^{-q\xi/2}}{q}.
    %V_\xi(q)=\frac{e^2}{2\epsilon}\frac{\tanh(q\xi/2)}{q}.
\end{equation}
We introduce $0\leq\lambda\leq1$ as a factor to suppress the intervalley interaction. This parametrization makes the interpretation of numerical results convenient, as it connects the physically relevant limit with full inter-valley interaction ($\lambda=1$) to known states in the decoupled regime, where the intervalley interaction vanishes ($\lambda=0$). The intra- and inter-valley interactions then take the form
\begin{align}
    V_{++}(q)=V_{--}(q)&=V_\xi(q),\\
    V_{+-}(q)&=\lambda V_\xi(q).
\end{align}
From this interaction potential, one can compute the Haldane pseudopotentials \cite{HPP1,HPP2,HPP3} for particles with relative angular momentum $m$ in the $n$th Landau level in the disk geometry via
\begin{equation}
    V_m^{(n)}=\int\frac{\textrm{d}^2q}{(2\pi)^2}V(q)[L_n(q^2\ell_B^2/2)]^2L_m(q^2\ell_B^2)e^{-q^2\ell_B^2},
    \label{eq:pseudopotential}
\end{equation}
where $L_m(x)$ are the Laguerre polynomials.
We now introduce the parameter $\delta V_0$ which tunes the onsite intervalley Coulomb term by modifying the Haldane pseudopotential $V_0^{(n)}\to V_0^{(n)}-\delta V_0$. Hence, $\delta V_0=0$ corresponds to the unmodified Coulomb interaction. We choose the normalization of the pseudopotentials such that $V_0^{(n)}=1$ and therefore the onsite intervalley repulsion exactly vanishes for $\delta V_0=1$. 
\begin{figure}
    \centering
    \includegraphics[width=\linewidth]{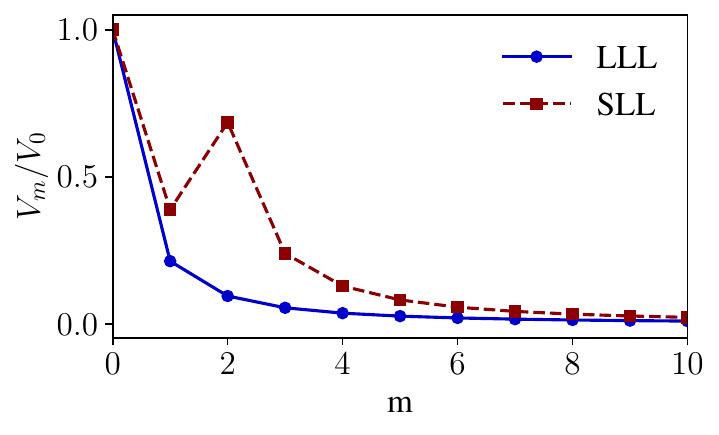}
    \caption{Pseudopotential coefficients $V_m^{(n)}$ for screened Coulomb interaction in the LLL with $n=0$ and SLL with $n=1$ as stated in Eq.~\eqref{eq:pseudopotential}.}
    \label{fig:pseudopotential}
\end{figure}

We work with an effective model for the SLL where we use the Haldane pseudopotentials of the SLL Coulomb interaction in the LLL. We do not explicitly use the SLL form factors. This approach allows us to perform calculations in the LLL while capturing the essential features of SLL physics \cite{Balram2020,Shi2008,bose2025dispersionneutralcollectivemodes}.
%Unless otherwise stated, the calculations are performed on the torus geometry.

\section{Results}

\begin{figure}
    \centering
    \includegraphics[width=\linewidth]{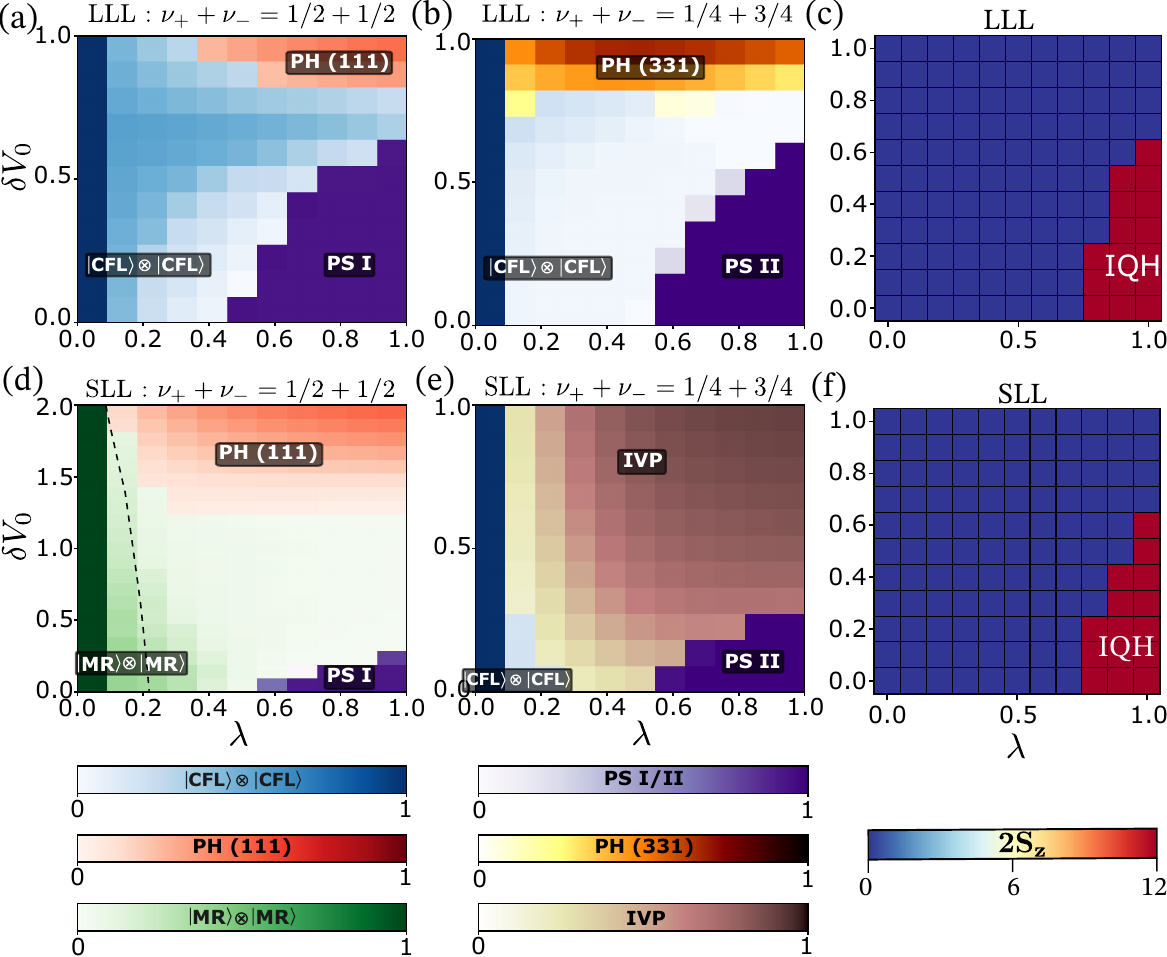}
    \caption{\textcolor{black}{Phase diagrams in terms of overlap and ground state polarization as a function of intervalley interaction strength $\lambda$ and the suppression of $V_0$. (a) phase diagram at LLL $\nu_++\nu_- = \frac{1}{2}+\frac{1}{2}$ considering $N_\phi=12$, (b) phase diagram at LLL $\nu_++\nu_- = \frac{1}{4}+\frac{3}{4}$ considering $N_\phi=16$, (c) polarization $2S_z = N_+ - N_-$ of the ground state in the LLL with $N_\phi=12$, (d)-(f) same as (a)-(c), but for SLL. Six colorbars correspond to the wavefunction overlap with the representative state listed in Tab.~\ref{tab:phases}. On the left side of the dashed line of (d) the  gap of the $|\text{MR}\rangle \otimes |\text{MR}\rangle$ state is finite.}}
    \label{fig:phase-diagram}
\end{figure}

Figure~\ref{fig:phase-diagram} shows the phase diagram in the LLL and SLL at the two filling factors $\nu=1/2+1/2$ and $\nu=1/4+3/4$. The shading in panels (a), (b), (d), and (e) indicates the overlap with six different trial wavefunctions, which we describe in detail below. In Fig.~\ref{fig:phase-diagram} (c), (f), we show the spin polarization of the ground state. We see that the ground state is either in the  $\nu=1/2+1/2$ or in the  $\nu=1+0$ sector. The only state in the $\nu=1+0$ sector is the spin-polarized integer quantum Hall state. One way to reach the partially spin-polarized case $\nu=1/4+3/4$ in the regime where the $\nu=1/2+1/2$ state is the ground state is by application of a finite magnetic field (see Supplement). We note, however, that differences between the ideal Landau level model and the realistic twisted MoTe$_2$ may lead to different ground state spin polarizations. One such difference is the finite band dispersion of twisted MoTe$_2$ which will have a tendency to depolarize the system with $\nu=1+0$. Hence, we nevertheless investigate the phases arising at $\nu=1/4+3/4$ in the full parameter range.

\begin{table}[ht]
    \centering
    \renewcommand{\arraystretch}{1.4}
    \begin{tabular}{c|c|c|c|c|c}
    \toprule
    phase & filling & LL & representative & GSD & gap? \\
    \midrule
    $|\text{CFL}\rangle \otimes |\text{CFL}\rangle$ & $\tfrac12+\tfrac12$ & LLL & $\delta V_0=0,\ \lambda=0$  & 144 & \xmark \\
    $|\text{CFL}\rangle \otimes |\text{CFL}\rangle$ & $\tfrac14+\tfrac34$ & LLL, SLL & $\delta V_0=0,\ \lambda=0$ & 144 & \xmark \\
    $|\text{MR}\rangle \otimes |\text{MR}\rangle$ & $\tfrac12+\tfrac12$ & SLL & $\delta V_0=0,\ \lambda=0$ & 36 & \cmark \\
    PS I & $\tfrac12+\tfrac12$ & LLL, SLL & $V_0=1$ & 36 & \xmark \\
    PS II & $\tfrac14+\tfrac34$ & LLL, SLL & $V_0=1$ & 128 & \xmark \\
    PH (111) & $\tfrac12+\tfrac12$ & LLL, SLL & $V_0=-1$ & 1 & \cmark \\
     &  &  &
      \multirow{3}{*}{%
        $\left\{\begin{array}{c}
           V_0=-1 \\
           V_1^{++}=1\\
           V_1^{--}=1
        \end{array}\right.$} &  &  \\
    PH (331) & $\tfrac14+\tfrac34$ & LLL &  & 8 & \cmark \\
     &  &  &  &  &  \\
    IVP & $\tfrac14+\tfrac34$ & SLL & $\delta V_0=10,\ \lambda=1$ & n/a & \xmark \\
    \bottomrule
    \end{tabular}
    \caption{\textcolor{black}{Phases arising in the phase diagram Fig.~\ref{fig:phase-diagram}. We list the degeneracies for $N_\phi=12$ for $\tfrac12+\tfrac12$ and $N_\phi=16$ for $\tfrac14+\tfrac34$.}}
    \label{tab:phases}
\end{table}

\subsection{Decoupled CFL}

    \begin{figure}
    \centering
    \includegraphics[width=\linewidth]{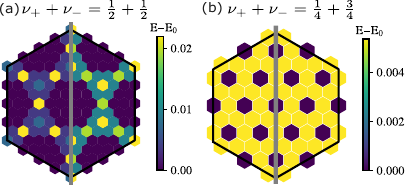}
    \caption{Comparison between composite fermion energy landscape (left side of the gray line in each figure) and exact energy landscape (right side of the gray line) for the CFL phase at LLL with $\lambda=0.0$, $\delta V_0=0.0$ for (a) $\nu=1/2+1/2$ at $N_\phi=12$, and for (b) $\nu=1/4+3/4$ at $N_\phi=16$. The left side of the gray line of each figure is for the CF model \cite{Fremling2018}, and the right side is from exact diagonalization. The energy bar is given for the exact diagonalization data. The total ground state degeneracy is 144 for both fractions. This shows good qualitative agreement between the CF energy and the exact energy for these two cases.}
    \label{fig:hexCFL}
    \end{figure}

Composite fermions are quasiparticles consisting of electrons bound to an even number of flux quanta. For the single valley case, at $\nu=1/2$ the composite fermions with two flux quanta attached to each electron experience no net magnetic field. Similarly, at $\nu=1/4$, the composite fermions with four fluxes attached experience no net magnetic field. In both cases, the composite fermions can then form a compressible state known as a CFL. Even though the state is gapless, in a finite-size system, there will be a gap separating the lowest energy states from the rest of the spectrum. The dispersion of the CFL has a characteristic form shown in Fig.~\ref{fig:hexCFL}. The ``degeneracy'' of the CFL and its spectrum can be understood from the dispersion of the composite fermions \cite{Fremling2018}. In the decoupled limit $\lambda=0$, the wavefunction is a product of two decoupled CFLs in the two valleys $\ket{\textrm{CFL}}\otimes \ket{\textrm{CFL}}$. The extent of the CFL in the phase diagram, being a gapless phase, is hard to pin down numerically. However, the characteristic low-energy landscape derived from decoupled composite fermions (Fig.~\ref{fig:hexCFL}) is observed to break down already for small intervalley coupling. We find the decoupled CFL states in the LLL at $\nu=1/2+1/2$ and in the LLL and SLL at $\nu=1/4+3/4$.

\subsection{Decoupled MR state}

    \begin{figure}
    \centering
    \includegraphics[width=\linewidth]{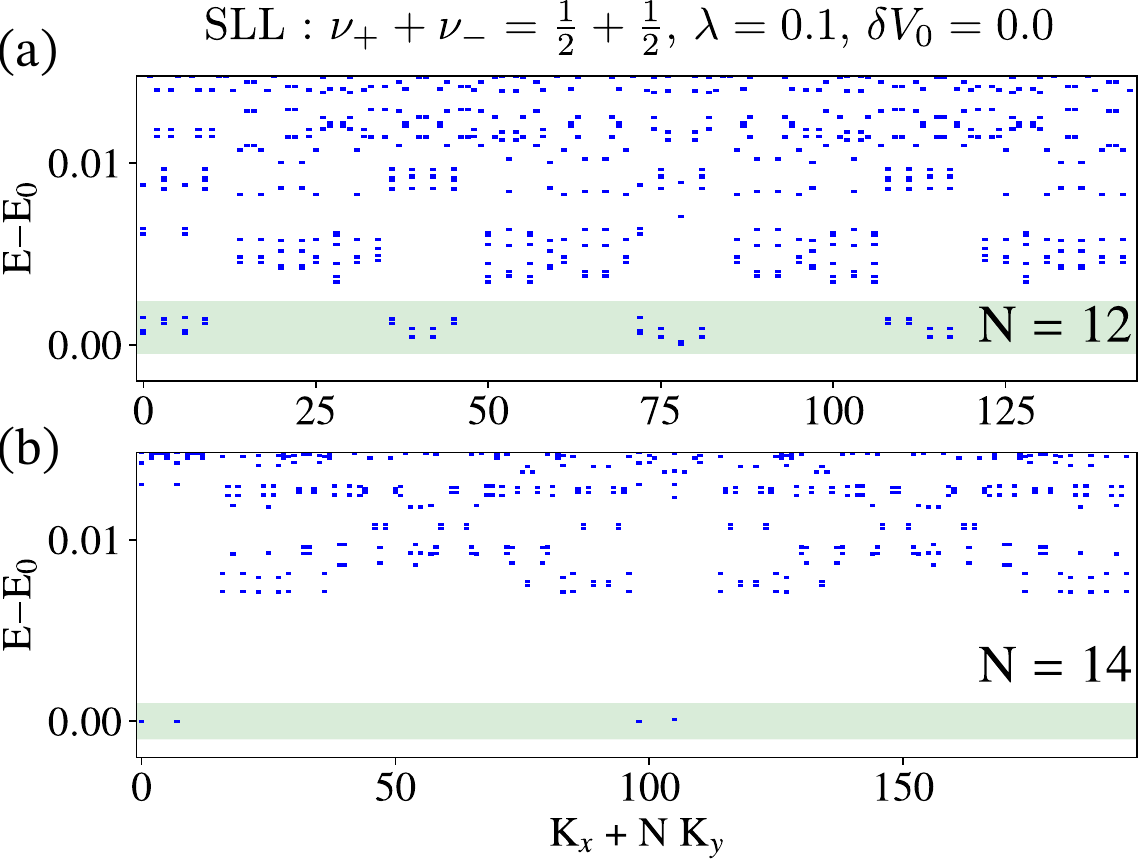}
    \caption{Energy spectra for two valley MR phase at SLL with $\lambda=0.1$ and $\delta V_0=0.0$ for (a) 12 particle system with 36-fold quasi-degenerate ground state, and (b) 14 particle system with 4-fold degenerate ground state as shaded by light green region. The ground-state degeneracy is consistent with the product of the MR state degeneracy for both odd and even particle systems.}
    \label{fig:MR}
    \end{figure}

While the CFL state is known to be the ground state of the Coulomb interaction at $\nu=1/2$ in the LLL, Fermi liquids can be unstable to pairing and this is what happens at $\nu=1/2$ in the SLL. The paired state of composite fermions is known as the MR state. This is a gapped state exhibiting non-Abelian topological order. The MR state is known to be the ground state in exact diagonalization studies of the half-filled SLL. This is relevant to the quantum Hall state occurring at filling $\nu=5/2$. The presence of the MR state can be checked by showing that the ideal MR state and the Coulomb ground state in the half-filled SLL are continuously connected \cite{Wang2009}.  The MR state has also been found in numerical studies of moiré systems~\cite{liu2024nonabelian,Chen2025,Wang2025,Xu2025}. At $\lambda=0$, the wavefunction is a product of two decoupled states in the two valleys. Each of the two decoupled states is in the same phase as the MR state, hence we can write the full state as  $\ket{\textrm{MR}}\otimes \ket{\textrm{MR}}$. Since the state is protected by a gap, the system remains in the same phase for a finite range of nonzero $\lambda$ around this decoupled point. This product of two quantum Hall states with opposite chirality is an example of a non-Abelian fractional topological insulator. 

One signature of the non-Abelian topological order of the MR state is the topological ground state degeneracy. In particular, for an odd particle number, the MR state is two-fold degenerate, while for an even particle number it is six-fold degenerate \cite{Read2000}. Taking the product of MR states with opposite chirality, we obtain 36-fold degeneracy when $N=4n$ and 4-fold degeneracy when $N=4n+2$. This degeneracy is exact at $\lambda=0$ and approximately persists at small finite $\lambda$, as can be seen in Fig.~\ref{fig:MR}.

Despite the stability for a finite range of $\lambda$, the gap closes for small values of $\lambda \simeq 0.2$, rendering this state an unlikely candidate for being realized in experiments, in which electrons experience isotropic density-density interactions $(\lambda=1)$.

%Finally, we note that there is a transition between the MR state and the Halperin 331 to MR as a function of interspin tunneling \cite{331_tunneling}.

\subsection{PS states}

    \begin{figure}
    \centering
    \includegraphics[width=\linewidth]{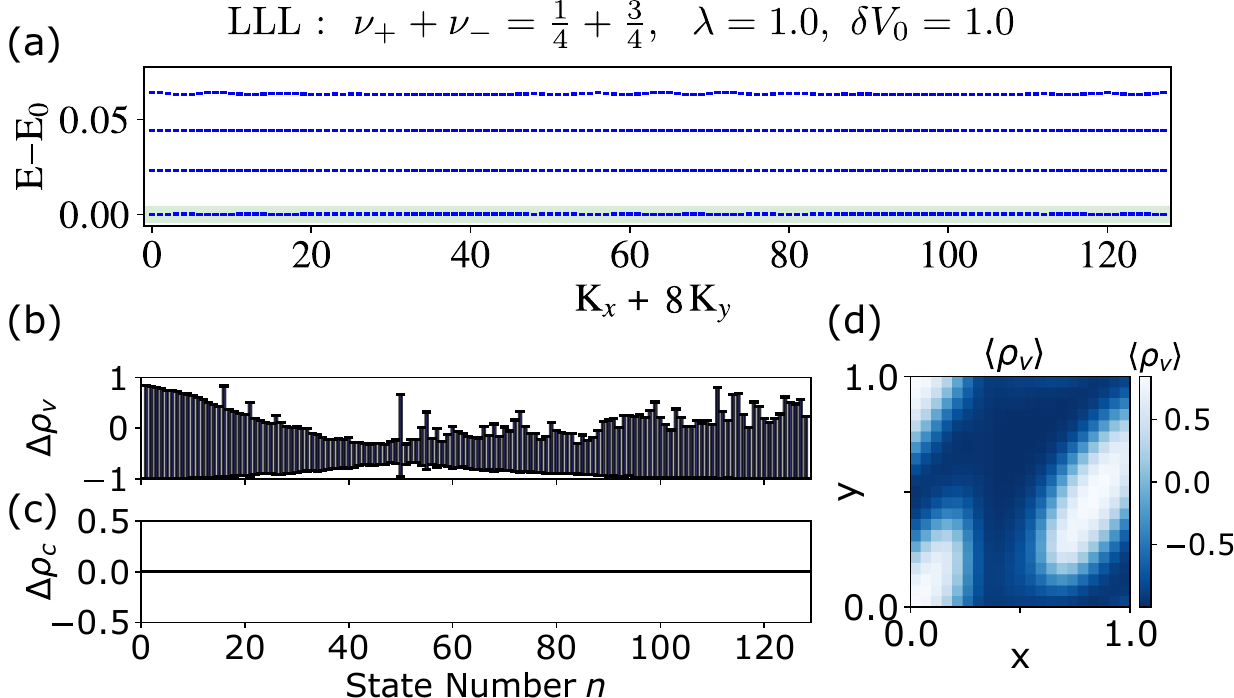}
    \caption{(a) Energy spectra for PS II state in the LLL with $\lambda=1.0$, $\delta V_0=0.0$ for $N_\phi=16$. All the momentum sectors contribute to the lowest energy state having total degeneracy $N.N_\phi$, (b)-(c) density fluctuation of valley ($\Delta\rho_v$) and charge ($\Delta\rho_c$) respectively for all 128 degenerate states, (d) valley polarization in the real space for one of the degenerate ground states. The valley polarization shows a large fluctuation while the charge density of the state remains uniform.}
    \label{fig:PS}
    \end{figure}

The phases adiabatically connected to a decoupled limit discussed so far do not persist upon the addition of sizable intervalley coupling. In the more widely studied situation of quantum Hall bilayers with equal Chern number, the electrons form a quantum Hall ferromagnet in the regime of strong layer coupling, thus minimizing the Coulomb repulsion. However, the quantum Hall ferromagnet is topologically obstructed in the case where the Chern numbers are opposite \cite{Bultinck2020mechanism,kwan2024textured}. Instead, one possibility to minimize the repulsive intervalley Coulomb interaction at fixed valley imbalance is a PS state, where the electrons in the two valleys spatially separate. This has also been seen in previous numerical calculations on the TVQH model~\cite{Mukherjee2019FQHsphereFTI,Chen2012FQHtorusFTI}.

Depending on the valley imbalance, $1/2+1/2$ or $1/4+3/4$, we get phases we call PS~I or PS~II, respectively.  The PS~I phase is $36$ fold degenerate with $N_\phi=12$, whereas the PS~II phase is $128$ fold degenerate with $N_\phi=16$. \textcolor{black}{For PS II there is one ground state in each momentum sector giving a total of $NN_\phi$ degenerate ground states. We have confirmed this for several system sizes.} The pseudopotentials we choose to generate reference states for overlap calculations in these two phases are the same ($V_0=-1$, all other pseudopotentials set to zero), however, the different valley imbalance results in a different ground state degeneracy. The spectrum of the PS~II phase shown in Fig.~\ref{fig:PS}~(a) has a ground state in every momentum sector.  We compute the valley and the charge fluctuations within the ground state manifold [see Fig.~\ref{fig:PS}~(b), (c)] and find that while the charge density of the state is uniform, there are large valley (and with it spin) polarization fluctuations (see Appendix B for details of the calculation). Charge inhomogeneities incur a Hartree energy penalty and are hence disfavored. On the other hand, valley polarization inhomogeneities are favorable due to the large onsite inter-valley repulsion $V_0$. We plot the valley polarization in real space for one example state in Fig.~\ref{fig:PS}~(d). The degenerate ground states can thus be understood as encoding clusters of valley-polarized electrons that can be translated in real space such that they can be centered in any of the unit cells of the system \cite{Crepel2024}.

At $\lambda=1$, we find from the valley polarization plots in Fig.~\ref{fig:phase-diagram} (c), (f), that the PS state is usurped by the spin polarized $\nu=1$ integer quantum Hall state, since this maximized the exchange energy of the system compared to the PS state. However, magnetic disorder may still favour domain formation.

The physical picture that emerges for these PS states is thus a system consisting of puddles with the same charge density but opposite valley polarization and Hall conductivities of $\sigma_{xy}= e^2/h$ and $\sigma_{xy}= -e^2/h$, averaging to (near) zero Hall conductivity in a macroscopic sample.

\subsection{Partially particle-hole transformed Halperin (331) state}

    \begin{figure}
    \centering
    \includegraphics[width=\linewidth]{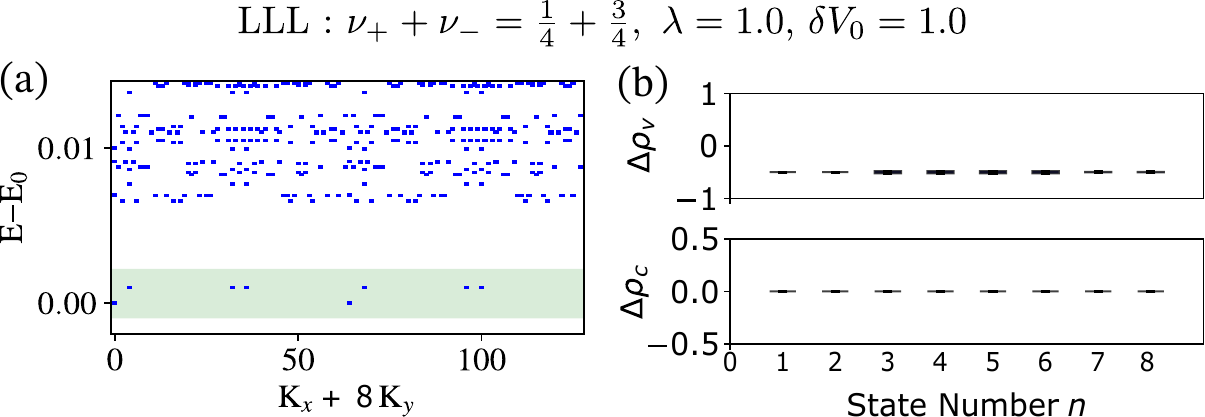}
    \caption{(a) Energy spectra for partially particle-hole transformed Halperin 331 state in the LLL with $\lambda=1.0,\ \delta V_0=1.0$, and $N_\phi=16$, (b) density fluctuation of valley ($\Delta\rho_v$) and charge ($\Delta\rho_c$) for the same phase considering 8-fold degenerate ground states.}
    \label{fig:Halperin331}
    \end{figure}

One family of states that have been used to describe equal Chern number quantum Hall bilayers are the Halperin $(mnl)$ states. They describe interlayer paired states at filling factors $\nu_+=\frac{1}{l+m}$ and $\nu_- =\frac{1}{l+n}$. In order to describe the TVQH case, one has to particle-hole transform one of the species. Such wavefunctions have been proposed in Ref.~\onlinecite{Sodemann}. In particular, the Halperin $(331)$ state describes a quantum Hall bilayer at filling factor $\nu=1/4+1/4$. After a particle-hole transformation of one species, it describes a TVQH state at $\nu=1/4+3/4$. Due to the partial particle-hole transformation, the wavefunction includes attractive correlations between the electrons in opposite valleys. It therefore only becomes relevant once the onsite intervalley repulsion has been sufficiently suppressed, i.e.,\ $\delta V_0$ needs to be sufficiently large. The Hall conductivity of the partially particle-hole transformed Halperin $(mml)$ state is \cite{Sodemann}
\begin{equation}
    \sigma_{xy}=\frac{e^2}{h}\bigg(\frac{2}{m-l}-1\bigg).
\end{equation}

The degeneracy of the Halperin $(mnl)$ state on the torus with $m,n>l$ is $mn-l^2$ \cite{Seidel2008}, hence the Halperin (331) state and its partially particle-hole transformed version have 8-fold degeneracy. This can indeed be observed in Fig.~\ref{fig:Halperin331}. We furthermore show that the state found in exact diagonalization has a uniform valley polarization and charge distribution, which is consistent with a Halperin state. Finally, in Fig.~\ref{fig:phase-diagram} (b), we show that the overlap of the exact diagonalization ground state with the partially particle-hole transformed Halperin (331) state is large.

\subsection{Partially particle-hole transformed Halperin (111) state}

    \begin{figure}
    \centering
    \includegraphics[width=\linewidth]{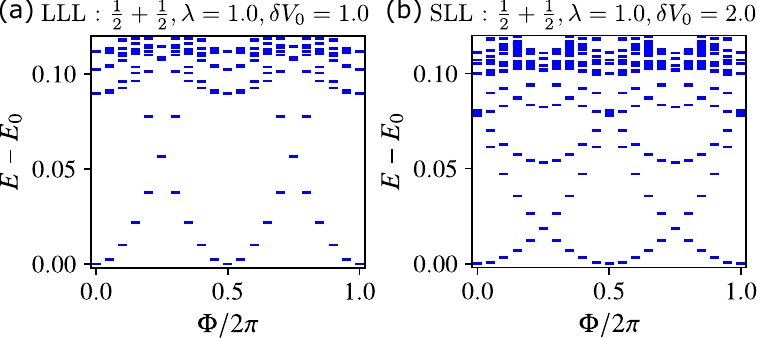}
    \caption{Flow of low energy states with flux insertion for partially particle-hole transformed Halperin (111) state at $\nu=1/2+1/2$ for (a) LLL with $\lambda=1.0$ and $\delta V_0=1.0$, and (b) SLL with $\lambda=1.0$ and $\delta V_0=2.0$. Both of these spectra show a periodicity $\pi$ under flux insertion.}
    \label{fig:SC}
    \end{figure}

The filling $\nu=1/2+1/2$ coincides with that of the partially particle-hole transformed Halperin (111) state. Indeed, we find good overlaps between this state and the exact diagonalization ground state in Fig.~\ref{fig:phase-diagram} (a), (d). The partially particle-hole transformed Halperin (111) state can alternatively be interpreted as a superconductor (SC) since the counterflow supercurrent of the Halperin (111) state in the quantum Hall bilayer setup becomes a charged supercurrent after the partial particle-hole transformation to go to the TVQH setup. This state occurs at a large (positive) $\delta V_0$ in both the LLL and SLL. Once the repulsive onsite interaction is sufficiently suppressed, pairing between electrons in opposite valleys can occur, leading to an SC. 

The transition between a fractional topological insulator and an SC has been studied in Ref.~\onlinecite{Crepel2024}. One important point is that for the usual quantum Hall setup, the SC is obstructed. If $c_\tau^\dagger(\bk)$ creates an electron with Chern number $C=\tau$, then $\langle c_+^\dagger(\bk)c_+^\dagger(\bk)\rangle$ needs to wind by $4\pi$ around the Brillouin zone. On the other hand, in the TVQH case we are studying, the SC is not obstructed, since $\langle c_+^\dagger(\bk)c_-^\dagger(\bk)\rangle$ does not wind around the Brillouin zone \cite{Li2018,Murakami2003}. Instead of pairing electrons, one could also consider pairing composite fermions. The SC obtained from the pairing of composite fermions in the TVQH model is topological \cite{zhang2018composite}.

The SC can be clearly identified numerically, since it is continuously connected to the strongly paired $V_0\to-\infty$ limit. As additional evidence, Fig.~\ref{fig:SC} shows that under flux insertion the spectrum has a periodicity of $\pi$, consistent with the presence of a charge $2e$ condensate \cite{Scalapino1993,Loder2008}.

%We note that signatures of unconventional SC have been observed in twisted MoTe$_2$ in proximity to the fractional Chern insulators \cite{xu2025signaturesunconventionalsuperconductivitynear}. 

\subsection{Intervalley paired (IVP) state}

    \begin{figure}
    \centering
    \includegraphics[width=\linewidth]{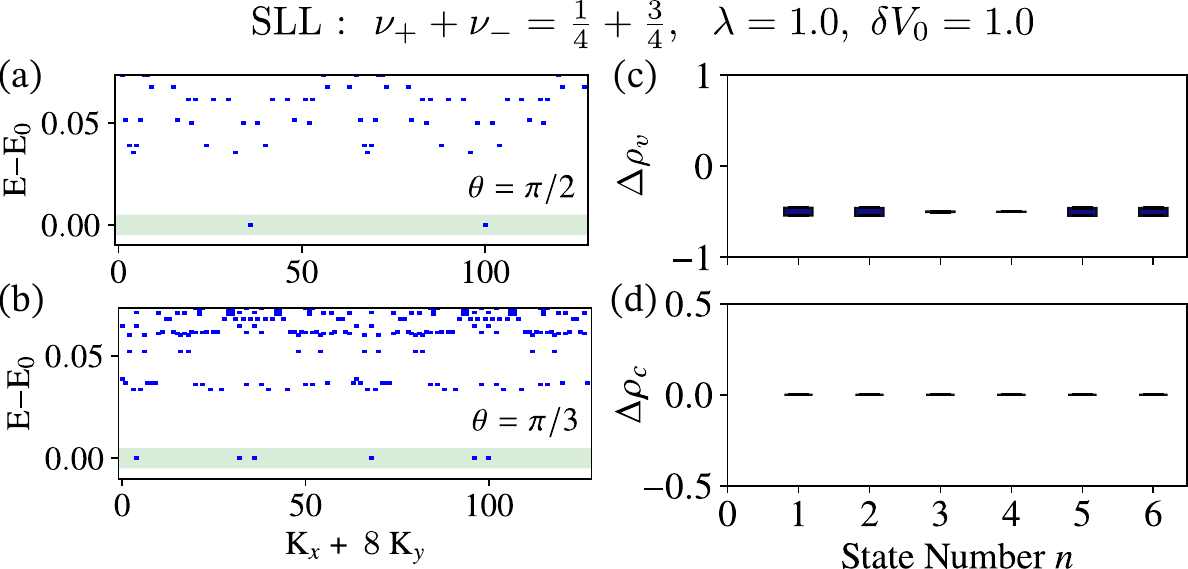}
    \caption{(a)-(b) Energy spectra for IVP state in the SLL with $\lambda=1.0,\ \delta V_0=1.0$, and $N_\phi=16$ for two different angles $\theta=\pi/2$ and $\theta=\pi/3$ respectively; (c)-(d) density fluctuation of valley ($\Delta\rho_v$) and charge ($\Delta\rho_c$) for the same state at $\theta=\pi/3$.}
    \label{fig:IVP}
    \end{figure}

We find a distinct state for a large (positive) $\delta V_0$ in the SLL, implying it is in the regime where intervalley pairing is favourable. The ground-state degeneracy in this regime depends on the angles between the axes of the torus as shown in Fig.~\ref{fig:IVP}: For $\theta=\pi/2$ \textcolor{black}{we have a square torus with $C_4$ symmetry and} the ground state is 2-fold degenerate. For $\theta=\pi/3$ \textcolor{black}{the Brillouin zone has $C_6$ symmetry and} the ground state is 6-fold degenerate. This dependence of the ground state degeneracy on the torus angle rules out a topologically ordered state. Furthermore, sweeping continuously the angle between $\theta=\pi/3$ and $\theta=\pi/2$, we see that the state has a gapless dispersion (see Supplement). We also see from Fig.~\ref{fig:IVP} that the charge and valley fluctuations of this state are small. 

 In the SLL, $V_2>V_1$ as shown in Fig.~\ref{fig:pseudopotential} and this is the pseudopotential which stabilizes this IVP state. In the Supplement, we show that this state can be very well described by taking the pseudopotential $V_2>0$, setting all other pseudopotentials to zero. This explains why the state is favoured in the SLL but not in the LLL.

\section{Connection to experiments in \titlelowercase{t}M\titlelowercase{o}T\titlelowercase{e}$_2$ at $\nu=-3$}

Experiments on twisted MoTe$_2$ at $\nu=-1$ typically see signatures of a spin-polarized integer Chern insulating state \cite{cai2023signatures,zeng2023integer,Park2023,Xu2023,park2024ferromagnetism,xu2024interplay}. On the other hand, experiments at $\nu=-3$ in a $2.1^\circ$ twisted MoTe$_2$ sample in a small magnetic field show a state with vanishing $\sigma_{xy}$, thus ruling out the obvious candidate at that filling, which is the spin-polarized integer Chern insulating state \cite{kang2024evidence_nature}. Despite the vanishing $\sigma_{xy}$, the state still breaks time-reversal symmetry, as can be seen from a finite magnetic circular dichroism signal \cite{kang2025timereversalsymmetrybreakingfractional}. Transport measurements reveal a state with finite (but not quantized) Hall conductivity at zero magnetic field, which is small compared to the quantum of Hall conductivity. With application of a small magnetic field, the Hall conductivity first gradually increases in magnitude, before entering a regime where it vanishes abruptly in finite field~\cite{kang2025timereversalsymmetrybreakingfractional}. \textcolor{black}{We note that $\sigma_{xy}$ has linear behaviour around zero and does not exhibit a plateau \cite{kang2024evidence_nature}.}

Out of the states we have found in our phase diagrams in the physical limit $\lambda=1$, the PS state could have a finite but non-quantized Hall response. On the other hand, it has been shown that the partially particle-hole transformed 331 state has a vanishing Hall response \cite{Sodemann}. It is therefore possible that in the absence of a magnetic field, the ground state is the spin-unpolarized PS state. Under application of a magnetic field, the spin-imbalanced Halperin 331 state could then be preferred, instead of a gradual flipping of domains, which would result in an integer quantum Hall state with a quantized Hall response. The latter is in contradiction to experiments. \textcolor{black}{Furthermore, since doping the partially particle-hole transformed 331 state results in itinerant charged particles which are hard to localize via disorder, this can explain the lack of a plateau \cite{Sodemann}.}

The way to avoid forming the integer quantum Hall state is to have a sufficiently large suppression of the onsite Coulomb repulsion $V_0$. Experimentally, the source of this suppression can be band mixing (i.e.,\ Landau level mixing in our Landau level model). The importance of band mixing has already been emphasized in previous work \cite{kwan2024abelianfractionaltopologicalinsulators}. Recent experiments see an SC in twisted MoTe$_2$ \cite{xu2025signaturesunconventionalsuperconductivitynear} hinting at the presence of an attractive interaction, which may be the source of the $\delta V_0$ used in our simulations to stabilize the 331 state. We note, however, that according to our calculations, the observation of an integer Chern insulator at $\nu=-1$ sets an upper bound on the suppression of the onsite repulsion $\delta V_0$. The spin gap of the IQH in the LLL at $\lambda=0.0,\ \delta V_0=0.0$ ($\simeq0.2142$) is larger than that in the SLL ($\simeq0.2066$), which indicates that the IQH is less robust in the SLL compared to the LLL.

The second valence band, which is the relevant band at $\nu=-3$, is in fact more similar to the SLL than the LLL \cite{Ahn_2024,Wang2025,Xu2025}. In the SLL, the 331 state does not appear, and instead, the ground state is the IVP state. Future work should confirm whether this state has a vanishing Hall conductivity and could hence also explain the experimentally observed state at small magnetic fields.

\section{Conclusion}

We study the TVQH model meant to capture the physics of twisted MoTe$_2$ at small twist angles of around $2.1^\circ$ at $\nu=-1$ or $\nu=-3$. We consider Landau levels with opposite Chern numbers, representing the spin-valley locked Chern bands in twisted MoTe$_2$. Due to its experimental relevance, we anticipate that the TVQH model may reach the same status as the bilayer quantum Hall model, which has been the subject of intense studies spanning decades during which numerical simulations have uncovered rich physics \cite{ED1,ED3,ED0,Park1,Park2,papicThesis,Wagner2021,Hu2024,Wagner2024,Simon2,Simon3,Milovanovic}.

Using exact diagonalization for a model with gate-screened Coulomb interactions, we identify a number of phases, some of which may be relevant for the interpretation of experimental results. 

For the case of fully spin-isotropic interactions, we find a PS state, a SC, a partially particle-hole transformed Halperin state, an integer quantum Hall state, and an IVP state. The tuning between these phases can be achieved by tuning the onsite Coulomb repulsion. Without modifying the Coulomb repulsion, the state would be an integer quantum Hall state. Reducing the onsite repulsion leads to a PS state, and eventually a paired state, once the onsite Coulomb repulsion is sufficiently softened to allow for pairing between opposite valley electrons. The paired state can take different forms depending on the spin polarization and on the precise form of the interaction potential. The paired state may be gapped, in which case it is adiabatically connected to a partially particle-hole transformed Halperin state. The robust partially particle-hole transformed Halperin (331) state with $\sigma_{xy}=0$ may explain the recently observed state at $\nu=-3$ in small magnetic fields \cite{kang2024evidence_nature,kang2025timereversalsymmetrybreakingfractional}. \textcolor{black}{Experimentally, it would be very interesting to measure the spin polarization at $\nu=-3$, since this would provide clear evidence for or against the partially particle-hole transformed Halperin (331) state.} Depending on the interaction potential, the paired state may also be gapless, leading to the IVP state we identify. Future work is required to identify the exact nature of the gapless IVP state. For the case of density-density interactions, we do not find a robust non-Abelian fractional topological insulator in our model, casting doubt on its relevance for the interpretation of the experiments of Refs.~\cite{kang2024evidence_nature,kang2025timereversalsymmetrybreakingfractional}. 
\\
\begin{acknowledgements}
We thank Nicolas Regnault, Yves Kwan, and Jiabin Yu for useful discussions.  
S.D. and T.N. acknowledge support from the Swiss National Science Foundation through a Consolidator Grant (iTQC, TMCG-2\_213805) and the Quantum Grant 20QU-1\_225225. 
G.W.~is supported by the Swiss National Science Foundation via Ambizione grant number PZ00P2-216183. Exact diagonalization calculations were performed using DiagHam.
\end{acknowledgements}

\begin{appendix}

\section{Pseudopotentials normalization}
    The pseudopotentials used for all the calculations in torus geometry have been taken from Eq.~\eqref{eq:pseudopotential} of the main text with a special normalization considering moiré lattice with twist angle $\theta_t$ which fixes the moiré momentum scale $$k_\theta=2|K| \sin{\theta_t/2}\,,$$ where $|K|=4\pi/\sqrt{3}a_0$, $a_0$ is the lattice constant. Now, the moiré reciprocal lattice vectors are
    $$\vec{b_1}=k_\theta (1,0), \quad \vec{b_2}=R_{120^\circ}\vec{b_1}=k_\theta\left(-\frac{1}{2}, -\frac{\sqrt{3}}{2}\right).$$
    $R_{120^\circ}$ symbolizes $120^\circ$ rotation. So, the total sample area of the $N_x\times N_y$ torus is
    $$A=N_xN_y\frac{(2\pi)^2}{|\vec{b_1}\times \vec{b_2}|}\,.$$
    We have normalized the pseudopotentials using the effective magnetic length $\ell_{B}=\sqrt{\text{A}/2\pi N_\phi}$ for a sample FCI having number of unit cells $N_x$ and $N_y$ so that the total flux $N_\phi=N_x.N_y$. For all of our calculations, we have taken twist angle $\theta_t=2.1^\circ$ and truncated the pseudopotentials up to $N_\phi$ values.

\section{Calculation of density fluctuations}

    To obtain the valley and charge density fluctuations, we first compute the valley-resolved one-body density matrix on the torus,
    $$g_{ij,\tau}(\vec{r}) \equiv \langle \Psi_i \,|\, \psi^\dagger_\tau(\vec{r})\,\psi_\tau(\vec{r})\,|\,\Psi_j\rangle, \quad \vec{r} =(x,y),\ \ \tau\in\{+,-\}. $$
    Expanding the field operator $\psi_\tau(\vec{r})$ in the single-particle torus orbitals $\phi_{n,k}(\vec{r})$,
    $ \psi_\tau(\vec{r})=\sum_{k=0}^{N_\phi-1}\phi_{n,k}(\vec{r}) c_{k,\tau} \, ,$
    gives the exact kernel form
    $$ g_{ij, \tau}(\vec{r}) = \sum_{k,k'=0}^{N_\phi-1}
    \phi_{n,k'}^*(\mathbf r)\,\phi_{n,k}(\vec{r})\,
    \langle \Psi_i|\,c^\dagger_{k',\tau}c_{k,\tau}\,| \Psi_j\rangle . $$
    The many-body ground states are expanded in Slater determinants,
    $$ |\Psi_i\rangle=\sum_l C^{(i)}_l\,|\Phi_l\rangle\, , $$
    where $C^{(i)}_l$ is the coefficient of Slater determinant $|\Phi_l\rangle$ built from the single-particle orbitals $\phi_{n,k}$ of the $n$th Landau level on a torus. We use
    \begin{align*}
        \phi_{n,k}(x,y) = & \mathcal{N}_n \sum_{m\in\mathbf{Z}} \exp{\left( \frac{2\pi}{L_y}(k+mN_\phi)(x+iy)\right)} \\ & \times\exp{ \left[-\frac{x^2}{2\ell_B^2}- \frac{1}{2}\left(\frac{2\pi\ell_B}{L_y}\right)^2 (k+mN_\phi)^2 \right] } \\ & \times H_n\left[ \frac{2\pi\ell_B}{L_y}(k+mN_\phi) - \frac{x}{\ell_B}\right].
    \end{align*}
    with normalization $\mathcal{N}_n$, magnetic length $\ell$, $n$th Hermite polynomial $H_n$, and $k=0,1,\dots,N_\phi-1$.

 From the above density matrices, we get,
    $$G_{+}(\vec{r}) = g_{ij,+}(\vec{r}) + g_{ij,-}(\vec{r}) ,\;\, G_{-}(\vec{r}) = g_{ij,+}(\vec{r}) - g_{ij,-}(\vec{r}).$$
    Then we calculate the eigenvectors, $v_i$, of any of the $G_+(\vec{r})$ or $G_-(\vec{r})$ at the origin ($\vec{r}=0$) for all degenerate ground states. In our calculations, we have considered $v_i$ for $G_+(\vec{r}=0)$. Now, due to the partial particle-hole transformation in our calculations, the density correlations for valley $\rho_s$ and charge $\rho_c$ can be calculated as
    $$\rho_{v}(\vec{r}) = (v_i^\dagger \cdot G_+(\vec{r}) \cdot v_i)/2N_e - 1\,, \;\, \rho_c(\vec{r}) = (v_i^\dagger \cdot G_-(\vec{r}) \cdot v_i)/2N_e$$
    respectively. Finally, the fluctuations of valley and charge for each degenerate state are 
    $$\Delta\rho_v = \text{max}(\rho_v) - \text{min}(\rho_v),  \quad  \Delta\rho_c = \text{max}(\rho_c) - \text{min}(\rho_c).$$
\end{appendix}

%\bibliography{refs}
\begingroup

\endgroup

\newpage
\clearpage
%\begin{appendix}
\onecolumngrid
	\begin{center}
		\textbf{\large --- Supplementary Material ---\\Fractional topological insulators at odd-integer filling: \\
Phase diagram of two-valley quantum Hall model}\\
		\medskip
		\text{Sahana Das, Glenn Wagner, Titus Neupert}
	\end{center}
	
		\setcounter{equation}{0}
	\setcounter{figure}{0}
	\setcounter{table}{0}
	\setcounter{page}{1}
    \setcounter{section}{0}
	\makeatletter
	\renewcommand{\theequation}{S\arabic{equation}}
	\renewcommand{\thefigure}{S\arabic{figure}}
	\renewcommand{\bibnumfmt}[1]{[S#1]}
    \renewcommand{\thesection}{S\arabic{section}}

\section{Additional figures}

    \begin{figure}[h]
    \centering
    \includegraphics[width=0.6\linewidth]{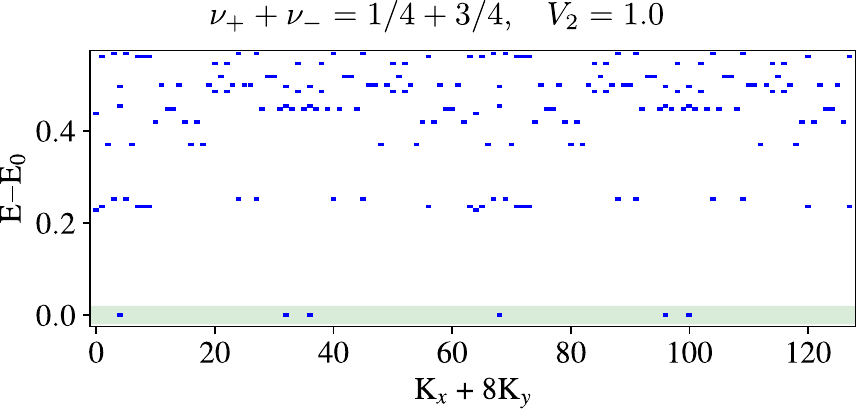}
    \caption{Energy spectra for $\frac{1}{4} + \frac{3}{4}$ with $N_\phi=16$ considering only non-zero inter-valley pseudopotential with relative angular momentum 2, $V_2$. In the SLL, $V_2$ is the most prominent pseudopotential after $V_0$. So, with the increase of $V_0$ suppression, $\delta V_0$, the emergence of the IVP state with 6-fold degeneracy is mostly taken care of by this prominent $V_2$ pseudopotential.}
    \label{fig:s1}
    \end{figure}

    \begin{figure}[h]
    \centering
    \includegraphics[width=\linewidth]{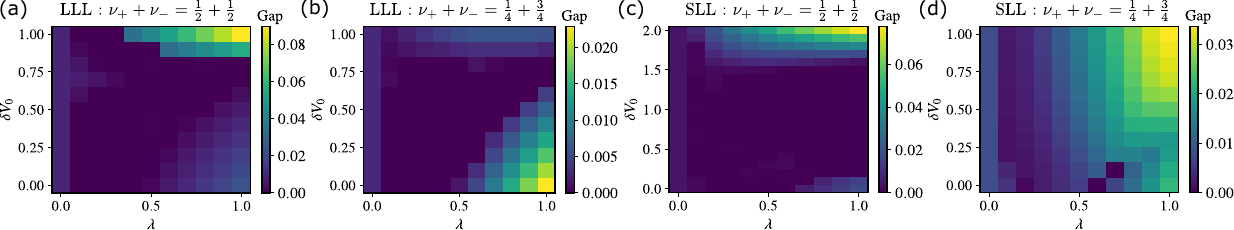}
    \caption{Energy gap as a function of intervalley interaction strength $\lambda$ and $\delta V_0$ for (a) LLL $\frac{1}{2}+\frac{1}{2}$ with $N_\phi=12$, (b) LLL $\frac{1}{4}+\frac{3}{4}$ with $N_\phi=16$, (c) SLL $\frac{1}{2}+\frac{1}{2}$ with $N_\phi=12$, and (a) SLL $\frac{1}{4}+\frac{3}{4}$ with $N_\phi=16$. The gaps are calculated considering the degeneracy corresponding to different phases of Fig.~\ref{fig:phase-diagram}. If $d$ is the degeneracy of any point according to Fig.~\ref{fig:phase-diagram}, then $\Delta_d$ will be the gap of that point.}
    \label{fig:s2}
    \end{figure}

    \begin{figure}
    \centering    
    \includegraphics[width=0.4\linewidth]{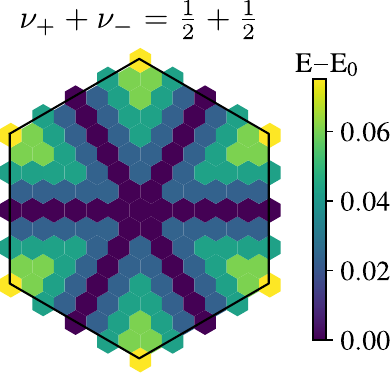}
    \caption{Hexagonal energy landscape for the PS I phase at LLL $\frac{1}{2} + \frac{1}{2}$ considering $N_\phi= 12$. The ground state degeneracy is 36.}
    \label{fig:s3}
    \end{figure}

    \begin{figure}
    \centering    
    \includegraphics[width=0.8\linewidth]{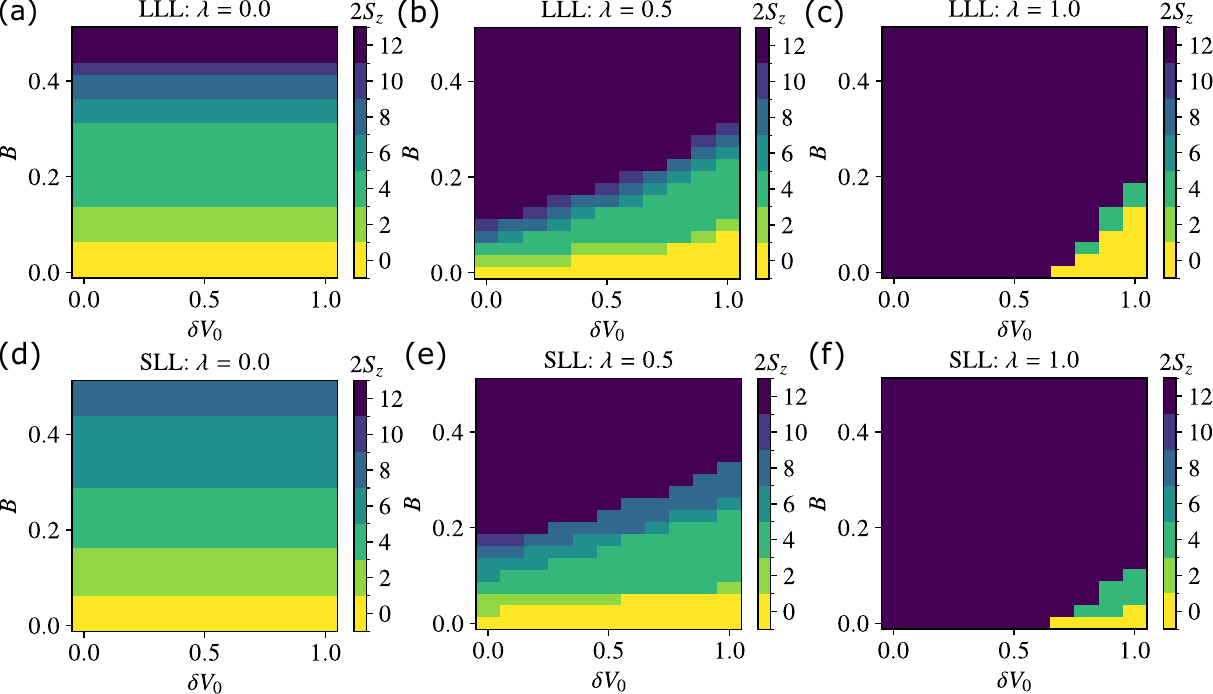}
    \caption{Magnetization landscape for various magnetic field B at $N_\phi=12$ where $E(B)= E-g\mu_B B S_z$ for (a)-(c) LLL with $\lambda=0.0,\ 0.5,\ 1.0$ respectively, and (d)-(f) same as (a)-(c) but for SLL. Here we have taken $ g=0.44,\, \mu_B=1$. In the absence of a magnetic field $B$, the ground state is either fully spin-polarized having $\nu=1+0$, or spin-unpolarized leading to $\nu=1/2+1/2$. But, with the application of a finite magnetic field, we can reach a partially spin-polarized state $\nu=1/4+3/4$.}
    \label{fig:s4}
    \end{figure}

    \begin{figure}
    \centering
    \includegraphics[width=0.5\linewidth]{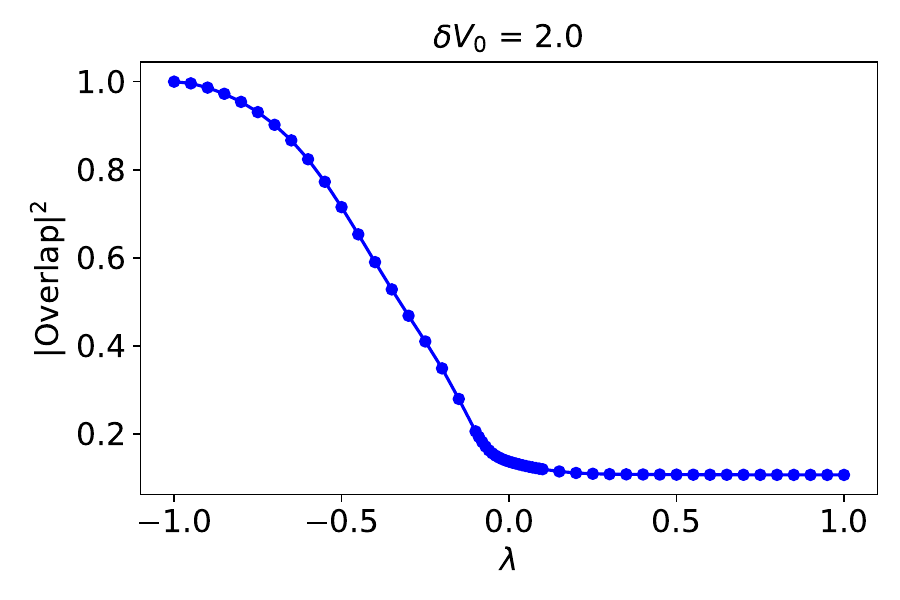}
    \caption{Overlap of ground states at SLL $1/4+3/4$, $\delta V_0 = 2.0$, and $N_\phi=16$ with the ground states at $\lambda=-1.0$. Due to the partial particle-hole transformation, $\lambda=-1.0$ refers to the repulsive Coulomb potential between two layers having equal weight of intra- and inter-layer interaction, which suggests it to be the MR state. The poor overlap value at the IVP state near $\lambda=1.0$ omits the possibility of this state becoming the MR state in spite of its same 6-fold ground state degeneracy.}
    \label{fig:s5}
    \end{figure}

    \begin{figure}
    \centering
    \includegraphics[width=0.8\linewidth]{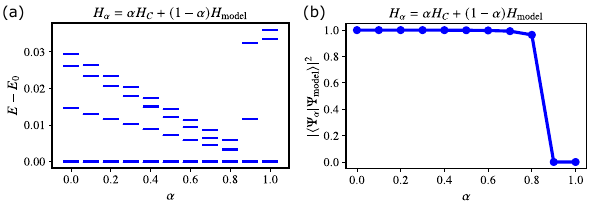}
    \caption{(a) Energy spectra of a Hamiltonian $H_\alpha$ as a function of $\alpha$, and (b) overlap of ground states of $H_\alpha$ with the ground states of a model Hamiltonian $H_\text{model}$.  $H_{C}$ takes pseudopotential values of the IVP phase in the second Landau level from Eq.~\eqref{eq:pseudopotential} with $\lambda=1.0$ and $\delta V_0=1.0$. $H_\text{model}$ is a model Hamiltonian with $V^1_{\tau,\tau}=1.0=-V^0_{\tau,-\tau}$, $V^3_{\tau,\tau}=0.4$ which corresponds to the inter-valley paired state of CFs described in Ref.~\cite{kwan2022hierarchy}. The gap closing of energy spectra near $\alpha\simeq0.8$ and sharp drop of overlap to zero near IVP phase omit the possibility of the IVP phase to become an inter-valley paired state of CFs.}
    \label{fig:s6}
    \end{figure}

    \begin{figure}
    \centering
    \includegraphics[width=0.6\linewidth]{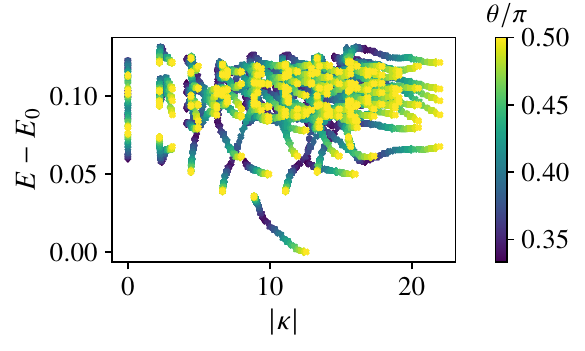}
    \caption{Energy spectrum as a function of $|\kappa|=2\pi \sqrt{r^{-1}k_x^2 + rk_y^2 - 2\cos{\theta}k_xk_y}/ \sqrt{N_e\sin{\theta}}$ at SLL $1/4+3/4$ with flux $N_\phi=16$ for $\lambda=1.0$ and $\delta V_0=1.0$, corresponding to the IVP phase. Spectra for different angles $\theta$ between $\pi/2$ and $\pi/3$ have been taken to interpolate between square ($\theta=\pi/2$) and hexagonal ($\theta=\pi/3$) cells. The aspect ratio $r$ is always 1. This clearly shows how the two-fold degenerate ground state at $\theta=\pi/2$ is connected to the six-fold degenerate ground state at $\theta=\pi/3$ leading to the conclusion that the state is gapless.}
    \label{fig:s7}
    \end{figure}

%\end{appendix}

\end{document}